\documentclass[%
twocolumn,
superscriptaddress,
 amsmath,amssymb,
 aps,
pra,
letterpaper,
]{revtex4-1}

\usepackage{amsmath,amssymb,amscd,amsxtra}
\usepackage{graphicx}
\usepackage{epstopdf}

\usepackage{color}

\newcommand\Rey{\mbox{\textit{Re}}}                               % Reynolds number
\newcommand\bu{\mathbf{u}}
\newcommand\sign{{\rm sign}}
\newcommand\hbu{\mathbf{\hat{u}}}

\newcommand\hp{\hat{p}}

\newcommand\br{\mathbf{r}}          % r vector
\newcommand\ex{\mathbf{e_x}}        % x unit vector
\newcommand\ey{\mathbf{e_y}}        % y unit vector
\newcommand\ez{\mathbf{e_z}}        % y unit vector
\newcommand\gx{\gamma_x}            % gammax
\newcommand\gy{\gamma_y}            % gammay
\newcommand\dxx{\delta_{xx}}        % deltaxx
\newcommand\dxy{\delta_{xy}}        % deltaxy
\newcommand\dyy{\delta_{yy}}        % deltayy
\newcommand\bx{\mathbf{x}}

\begin{document}

\title{Direct measurement of particle inertial migration in rectangular microchannels}

\author{Kaitlyn Hood}
 \email{kaitlyn.t.hood@gmail.com}
\affiliation{%
 Department of Mathematics,
 University of California Los Angeles,
 Los Angeles, CA 90095, USA
}%

\author{Soroush Kahkeshani}
\affiliation{
 Department of Bioengineering,
 University of California Los Angeles,
 Los Angeles, CA 90095, USA
}

\author{Dino DiCarlo}
\affiliation{
 Department of Bioengineering,
 University of California Los Angeles,
 Los Angeles, CA 90095, USA
}

\author{Marcus Roper}
\affiliation{%
 Department of Mathematics,
 University of California Los Angeles,
 Los Angeles, CA 90095, USA
}%

%\date{\today}

\begin{abstract}
      Particles traveling at high velocities through microfluidic channels migrate from their starting streamlines due to inertial lift forces.  Theories  predict different scaling laws for these forces and there is little experimental evidence by which to validate theory.  Here we experimentally measure the three dimensional positions and migration velocities of particles.  Our experimental method relies on a combination of sub-pixel accurate particle tracking and velocimetric reconstruction of the depth dimension to track thousands of individual particles in three dimensions. We show that there is no simple scaling of inertial forces upon particle size, but that migration velocities agree well with numerical simulations and with a two-term asymptotic theory that contains no unmeasured parameters.
\end{abstract}

\maketitle

%%%%FONT SETUP - please do not change any commands within this section
%\renewcommand*\rmdefault{bch}\normalfont\upshape
%\rmfamily
%\section*{}
%\vspace{-1cm}

%%%%FOOTNOTES%%%
%
%\footnotetext{\textit{$^{a}$~Department of Mathematics, UCLA, Los Angeles, CA, USA.}}
%\footnotetext{\textit{$^{b}$~Department of Bioengineering, UCLA, Los Angeles, CA, USA.}}
%
%%Please use \dag to cite the ESI in the main text of the article.
%%If you article does not have ESI please remove the the \dag symbol from the title and the footnotetext below.
%\footnotetext{\dag~Electronic Supplementary Information (ESI) available: [details of any supplementary information available should be included here]. See DOI: 10.1039/b000000x/}
%%additional addresses can be cited as above using the lower-case letters, c, d, e... If all authors are from the same address, no letter is required
%
%\footnotetext{\ddag~Email: ektuley@math.ucla.edu}
%
%
%%%%END OF FOOTNOTES%%%

%%%MAIN TEXT%%%%

\section*{Introduction}

Inertial migration; the systematic movement of particles across streamlines due to finite Reynolds number forces, is exploited in systems to separate, focus and filter particles and cells \cite{DiCarlo14}. Though there are many theories for the magnitudes of inertial focusing forces, direct experimental measurement of these forces remains an unmet challenge. Indeed existing theory \cite{Saffman65,HoLeal74, Hinch89, Hood15}, numerical simulations \cite{DiCarlo09, ProhmStark14, Liu2015}, and indirect experimental measurements \cite{Papautsky13} have produced contradictory scalings for the dependence of forces on particle size and velocity.  In this paper, we directly measure inertial migration velocities by tracking the motion of particles in a rectangular channel over Reynolds numbers ranging from 30 to 180, and find that their measured migration velocities agree well with existing asymptotic theory \cite{Hood15}.

\begin{figure*}
    \includegraphics[scale=.5]{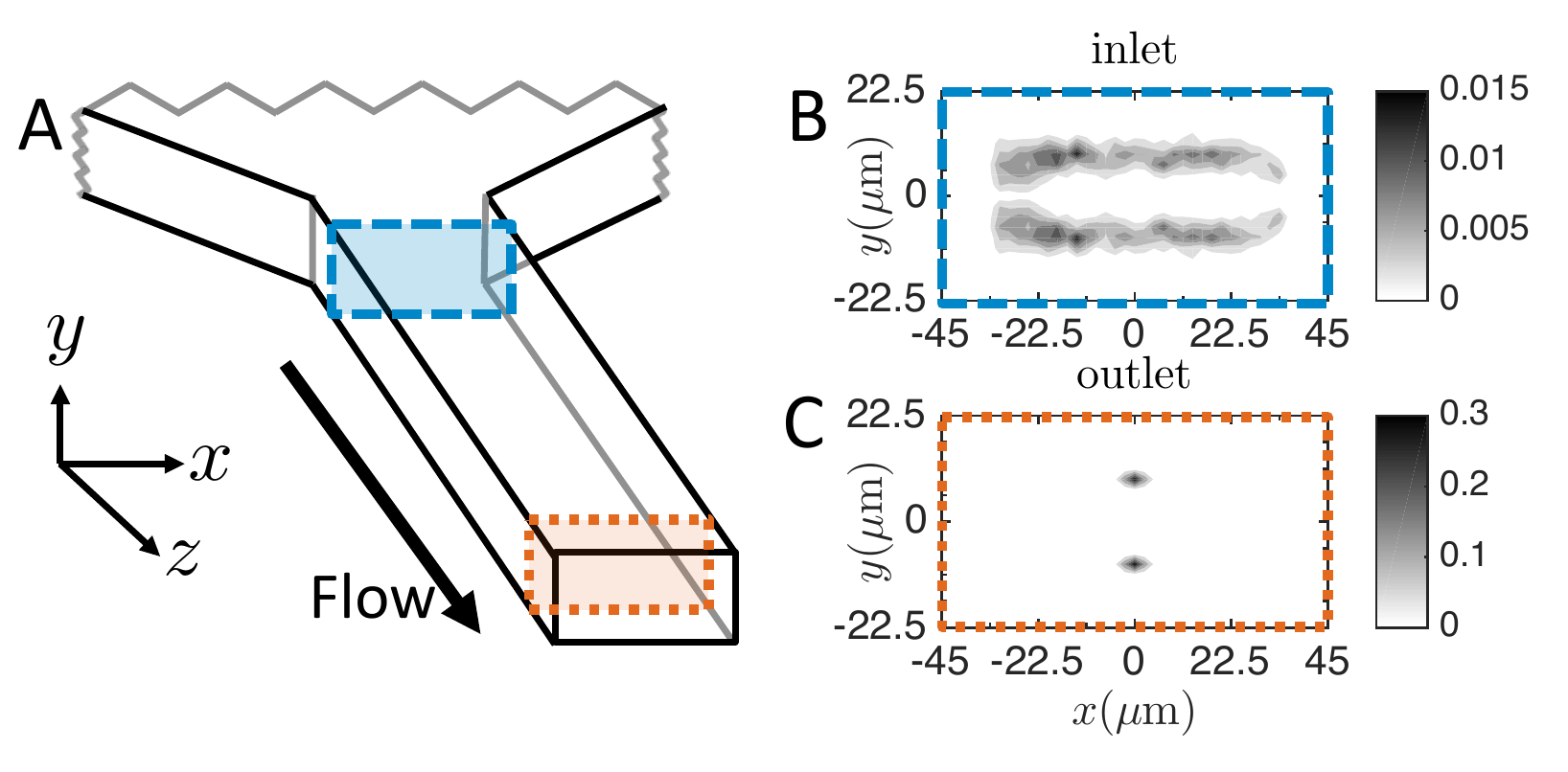}
    \includegraphics[scale=.5]{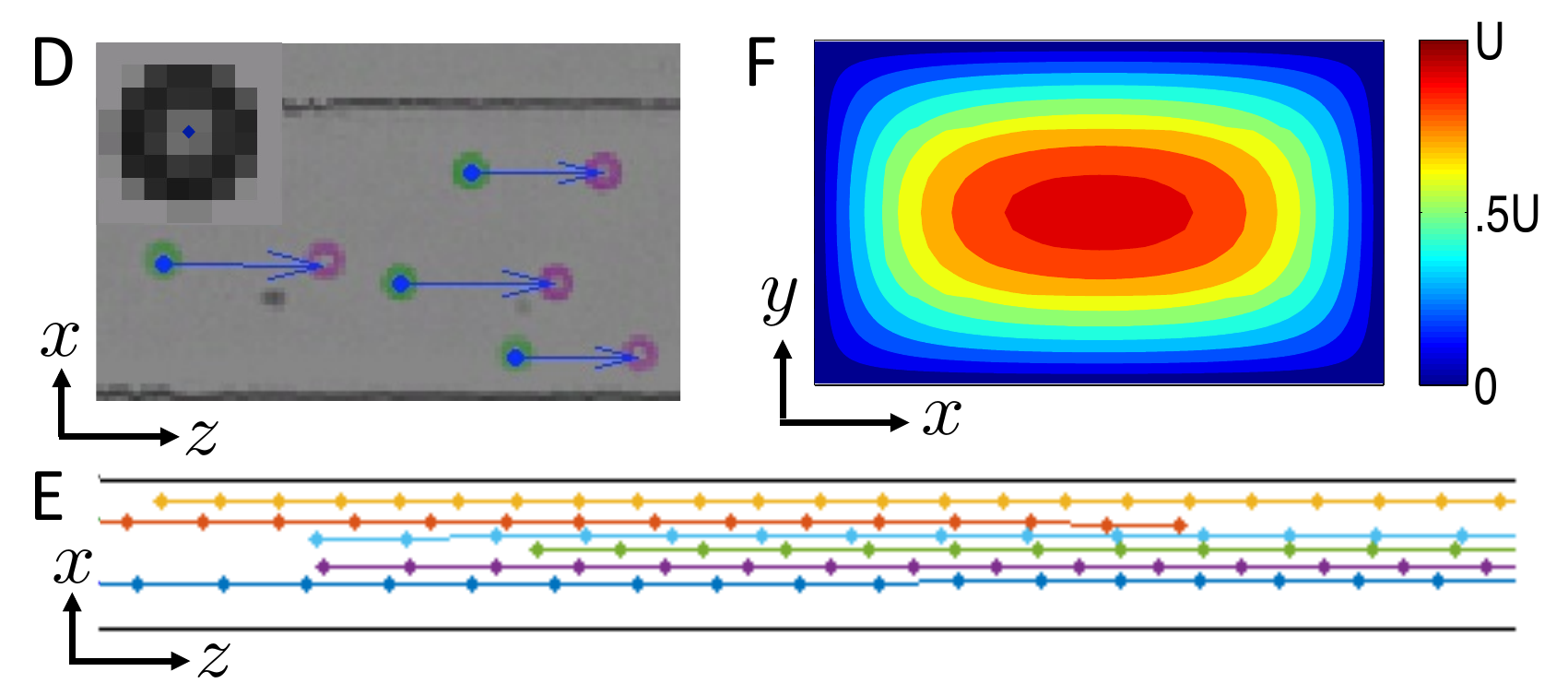}
   \caption{\label{fig:channel} Reconstruction of particle focusing velocities and three dimensional positions in a rectangular channel. (A) Schematic of the inlet of the channel. (B) Reconstructed probability density function (PDF) of particle distributions across the channel cross-section for 10\,$\mu$m particles at $Re=30$ shows that within the first 1 mm of the channel particles are initially focused to two narrow bands of streamlines (density shown in grayscale). (C) After 1.5 cm of inertial focusing, the same particles are fully focused to two streamlines on the channel mid-line.  (D) A hybrid PIV-particle tracking scheme is used to track the particles, green circles show particles in present frame, magenta circles show the particles in the next frame. (inset) Template matching (blue circle) allows particle center to be located with sub-pixel accuracy. (E) Representative trajectories of six particles tracked over 700\,$\mu$s.  (F) Numerically computed downstream particle velocity as a function of $x$ and $y$ positions across the channel cross-section: using this plot and the particle velocity in the $z$-direction, we can compute its $y-$position.}
\end{figure*}

Inertial migration of neutrally buoyant particles was first reported in flows through circular pipes \cite{SegreSilberberg61}. In a pipe with radius $R$, particles are inertially focused into a ring with radius approximately $0.6R$.  Furthermore, particles with different sizes are focused at different rates and to rings with slightly different radii \cite{JeffreyPearson65, Karnis66, Tachibana73, Uijttewaal94, ChoiLee2010}.  However,  microfluidic channels are more readily built with a rectangular geometry, in which particles are inertially focused to either two or four stable equilibrium streamlines \cite{DiCarlo09}. Focusing occurs in two phases, with apparently well-separated natural time scales:  (Fast phase) first particles quickly focus to a two-dimensional manifold of streamlines and then (Slow phase) particles travel within the manifold to one of the focusing streamlines. Two stage focusing has been experimentally measured \cite{ChoiSeoLee11}, and is consistent with numerical simulations of the spatial pattern of lift forces across the channel cross-section \cite{SMLL:DiCarlo2012,ProhmStark14,Liu2015,Hood15}.

Focusing in rectangular channels has been studied asymptotically, generating disagreement over the scaling of the inertial lift force. Recent asymptotic studies \cite{HoLeal74, Hinch89, Asmolov99} predicted that the inertial lift force $F_L$ is proportional to the fourth power of the particle radius $a$, i.e. $F_L \sim a^4$. This scaling hinges on the assumption that the particle radius is asymptotically smaller than the channel size, $a \ll H$.  Di Carlo et al showed that the $a^4$ scaling did not agree with numerical simulations \cite{DiCarlo09}.  Rather,  the numerical data suggested $F_L \sim a^3$.  Hood et al \cite{Hood15} extended the asymptotic analysis of Ho \& Leal \cite{HoLeal74}.  The resulting scaling law $F_L \sim c_4 a^4 + c_5 a^5$, reconciles the asymptotic scaling $F_L \sim a^4$ in the limit $a\ll H$ with the numerical data of Di Carlo et al \cite{DiCarlo09} up to experimentally used particle sizes, in which $a \sim H$. By contrast, Saffman's asymptotic study of inertial lift force assumes that the particle experiences an external force in the direction of flow in addition to the inertial lift force \cite{Saffman65}.  Using an indirect experimental measurement of inertial focusing, Zhou and Papautsky \cite{Papautsky13} report that $F_L \sim a^2$, in agreement with Saffman.  But they do not explain why Saffman's result applies to particles that are traveling freely with the flow of fluid.

Here we present the first reconciliation of predictive theory and direct experimental measurement of inertial migration velocities. While holographic techniques have been used to measure 3D particle distributions and velocities in microfluidic capillaries \cite{sheng2008using, katz2010applications, ChoiSeoLee11, choi2012advances}, but to the best of our knowledge holographic techniques have not been used to measure inertial migration velocities.  In this paper we propose an alternative to holographic techniques for measuring the 3D positions and velocities in PDMS microchannels. Our method allows accurate measurement of particle migration velocities in two dimensions, and via a velocity-based reconstruction method, of their position in the third dimension. This method provides position readouts for thousands of particles and allows particle positions and particle trajectories to be measured.  Thus, our method provides the first direct measurement of inertial migration velocities.  In addition to verifying the existence of a slow-focusing manifold, our position measurements show that significant inertial focusing occurs while particles are funneled into the channel, and that once this contribution is accounted for, inertial migration velocities agree fully with an asymptotic theory \cite{Hood15}.

\section*{Experimental methods}

Inertial focusing was measured in a 1.5 cm long PDMS microchannel fabricated using Sylgard 184 PDMS kit (Dow Corning Corp.) bonded to a glass slide as shown in Duffy et al \cite{Whitesides98}. The microchannel mold was fabricated using KMPR 1025 (MicroChem). The channel cross-section dimensions were $90 \mu{\rm m} \times 45 \mu{\rm m}$ $(W \times H)$, respectively, with the shortest dimension identified as the depth ($y$) dimension (Fig. 1A) and the longer dimension as the width or lateral dimension ($x$). The schematics of the channel are displayed in Fig. \ref{fig:channel}A. Particles enter the channel through an contracting inlet region whose depth is constant ($45\mu$m) and tapers in width from 1.5 mm to 90 $\mu$m over a 2.4 mm downstream length.  

The particles were dispersed at 0.004 volume fraction in a suspending fluid composed of deionized water and 0.002 (wt/vol) triton X-100. This suspension was pumped into the channel at controlled flow rate using a syringe pump (Harvard Apparatus, Holliston MA).  The solutions were infused using PEEK tubing (Idex: 1/32'' OD*0.02'' ID*5ft). The polystyrene spherical particles were chosen to be near-neutrally buoyant with a particle density of 1.05 g/cm$^3$. The particle density does not match the density of the suspending fluid (density 1.00g/cm$^3$), never the less the effects of sedimentation can be ignored in this experiment.  The sedimentation velocity can be determined by balancing buoyancy force with the drag force for a sphere.  For this experiment, the sedimentation velocity is at most 10$\mu$m/s, meaning that the particles sediment a distance of less than 0.3 $\mu$m over the entire length of the channel.  Therefore, sedimentation effects are negligible compared to the downstream velocity ($\sim$0.6m/s) and inertial migration velocity ($\sim$3mm/s).  

The channel Reynolds number is defined by $\Rey = U H/ \nu$, where $\nu = 1\times10^{-6}$ m$^2$/s is the kinematic viscosity of deionized water at room temperature, $H=45\mu$m is the short dimension of the channel, and $U$ is the average fluid velocity in the channel.  The ratio of particle size to channel size is defined by $\alpha = a/H$, where $a$ is the particle radius, and the particle Reynolds number is given by $\Rey_p = \alpha^2 \Rey = U a^2 / \nu H$.  Four particle radii were separately used, $a = 2.4$, $5$, $6$, and $9.5 \mu$m, along with four different total flow rates $Q = 160$, $320$, $640$, and $960 \mu$L/min, corresponding to a range of channel Reynolds numbers $Re=30-180$ and particle Reynolds numbers $Re_p=0.08-3.2$. The maximum Reynolds number of 180 was chosen to avoid delamination of the PDMS from the glass slide, while the minimum Reynolds number of 30 was chosen so that the inertial particle migration rate would be observable in the channel of length of 1.5cm.

Particle velocities were tracked by high speed imaging (14000 frames per second and 2$\mu$s exposure time, using a Phantom V710 camera) over the first and last $1$ mm of the channel. The microchannel was viewed from above using a microscope (Nikon Ti-U) with 4x objective with effective pixel size of 3$\mu$m.  The depth of field is listed to be 50 $\mu$m by the manufacturer, however blurry particles are still observable even for a range of upwards of 200 $\mu$m, so that the particles can be observed over the entire channel depth. For all diameters and velocities, particles were eventually focused to two streamlines on the mid-plane $x=0$ (Fig. \ref{fig:channel}B-C).

\section*{Determining the particle migration velocity}

High speed videography provided only $x$- and $z$- (lateral and streamwise) coordinates for each particle, and provided no direct measurement of the particle depth ($y$-coordinate). We measured the $x$- and $z$- velocities by hybridizing particle image velocimetry (PIV) and particle tracking, similar to an algorithm previously developed for tracking fluorescent organelles \cite{Roper13}.  First, we use the PIV code MatPIV \cite{matpiv} to develop a vector field representing the displacements of all particles from one frame to the next. Second, template matching is used to align a template consisting of a single 8$\times$8 pixel image of a particle with both the first frame and the next. The template matching process gives a single correlation value for every pixel in the image, representing how closely the template matches the real image centered at that pixel. Then we use cubic polynomials to interpolate the correlation data and find each particle location with sub-pixel precision. After locating particles in both frames, the PIV velocity field is used to predict the particles' locations in the subsequent frame. We identify the detected particle in the next frame that is closest to this predicted location. The particle tracking adjustment allows us to correct PIV velocity fields to obtain sub-pixel accurate particle displacements (Fig. \ref{fig:channel}D).

Multiple frames are needed to measure the migration velocity since the lateral displacements of particles over a single frame are typically sub-pixel.  Indeed, inertial migration velocities are typically two orders of magnitude smaller than particle downstream velocities (3 mm/s in a typical experiment compared to 0.6 m/s downstream velocity). To accurately measure the migration velocities, we track single particles over at least 10 consecutive frames, and average their total lateral displacement over all of these frames (Fig. \ref{fig:channel}E).

We reconstruct the $y-$positions of the particles using a numerical prediction of the downstream velocity.  We used a finite-element model built in Comsol Multiphysics (Comsol, Los Angeles) to compute the downstream velocities for force-free and torque-free finite particles whose size matched the experiments \cite{Hood15} located anywhere within the channel (Fig. \ref{fig:channel}F).  The Stokes timescale $\tau_s = 2\rho a^2/9\mu$, gives a measure of the time needed for a particle at any point in the channel cross-section to accelerate until it is both force and torque free. For the particles in our study $\tau_s=5-80\,\mu$s, is much less than a typical tracking time of $700\,\mu$s, so particles are effectively force-free and torque-free throughout their migration.  Downstream velocities vary across the depth of the channel, with no slip boundary conditions on the upper and lower walls of the channel and fastest velocities attained on the mid-plane of the channel.   For each $x$-position there is a two-to-one mapping of downstream velocity to particle depth, allowing particles to be assigned one of two $y-$coordinates that are symmetric about the depth mid-plane $y=0$ (Fig. \ref{fig:channel}F).

We measured the two dimensional probability density function (PDF) for the $x-$ and $y-$ coordinates of particles at the entrance to the microchannel and after 1.5cm of inertial focusing (Fig. \ref{fig:channel}B-C).  Particles within 1mm of the microchannel entrance are not uniformly dispersed in channel depth but instead are focused to a thin band of $y-$ coordinates (Fig. \ref{fig:channel}B).  We call this phenomenon pre-focusing because it is a consequence of inertial migration that occurs in the contracted inlet region before the particle enters the channel.  Along the channel, particles move laterally within this band until they are also focused close to the channel center-line, with typically 71\% of particles focused to within 4\,$\mu$m of the focusing streamline after traveling 1.5cm through the microchannel (Fig. \ref{fig:channel}C). 

The thin band on which particles are concentrated in the first 1 mm of the channel coincides with an asymptotic calculation for the slow manifold, described in more detail below (Fig. \ref{fig:piv_manifold}A-D). Since the particles are already focused to their slow manifold, the observed lateral migration within the microchannel represents only the second phase of particle focusing, i.e. the migration of particles along the slow manifold to their eventual focusing streamline (Fig. \ref{fig:piv_vel}). %The small discrepancy could be a result of deformations of the PDMS channel up to 10\% of the channel width $W$ \cite{sollier2011rapid}. 

\section*{Validation of the reconstruction algorithm}

In order to validate the measurement of particle heights via the velocimetric method, we ran the following experiment to independently measure the particle heights. Since particles outside the focal plane appear blurry, we exploit this blurriness to distinguish particle heights.  We will call this method the laplacian algorithm, because it uses the discrete Laplacian to measure the sharpness of the edges of the particle.

The experiment is designed as follows: we vary the focal plane height of the microscope and at each height measure the number of particles that appear to be in-focus. In this experiment there are two potential sources of blur: out-of-focus blur and motion blur.  In order to reduce the motion blur, we ran this experiment at $\Rey = 1$ and flow rate $Q = 5\mu$L/min.  We used $12\mu$m diameter particles and kept the exposure time constant ($2\mu$s) and reduced the frame rate to 500fps. During the experiment the focal plane is raised in $6\mu$m increments.  We measured these increments using a Nikon inverted microscope with programmable focus, which allows the focal plane to be precisely controlled. 

The laplacian algorithm works as follows.  We average the discrete Laplacian on a $7 \times 7$ pixel sub-image around the particle to get a single laplacian measurement for each particle.  A larger value indicates the particle is more in focus, and we can reference each value against a calibration measurement to measure the relative height of the particle to the focal plane.  The calibration measurement comes from running the laplacian algorithm on stationary particles resting on the bottom of the channel at various focal plane heights.

At each focal plane height we count the number of particles that are measured to be within $3\mu$m of the focal plane via both the laplacian and velocimetric algorithms.  Recall that the reconstruction algorithm cannot distinguish between particles in the top half of the channel and the bottom half, so we use the laplacian algorithm to make that distinction.

\begin{figure}
    \centering
    \includegraphics[scale=.45]{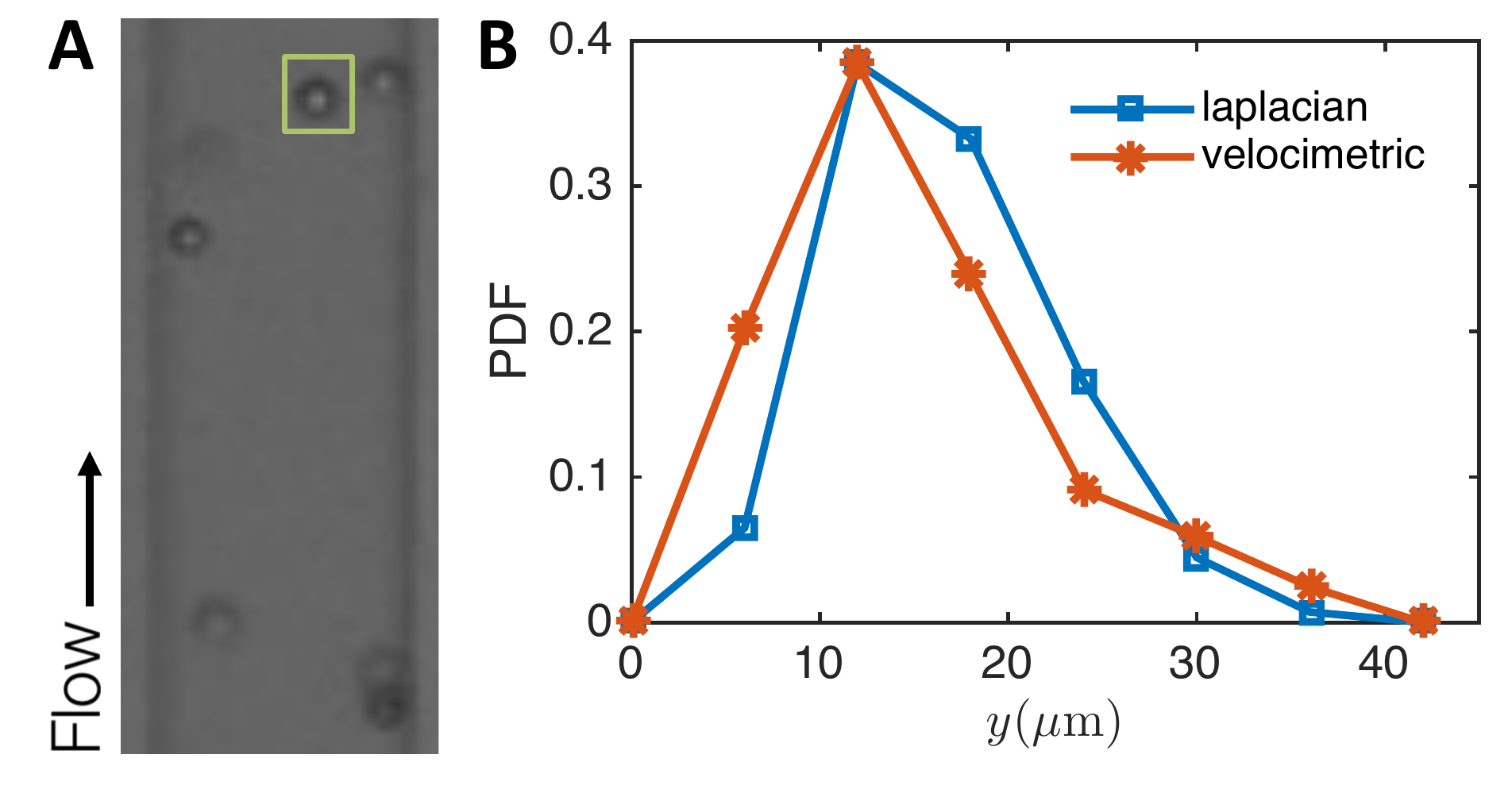}
   \caption{\label{fig:laplacian} Validation measurement for $\Rey = 1$ and $a = 6\mu$m. (A) Raw image of particles with focal plane height of $6\mu$m above the bottom of the channel.  The laplacian algorithm measures only one particle to be in focus (inside green box).   (B) Experimental calibration of the particle height measurement.  The PDF of particle height compares well between the velocimetric algorithm (blue square) and the laplacian algorithm (orange asterisk).}
\end{figure} 

A comparison of the PDF of particles via the velocimetric algorithm and particles via the laplacian algorithm shows good agreement (Figure \ref{fig:laplacian}).  We observe that the particles are much more likely to be in the bottom half of the channel.  This is to be expected, since at $\Rey = 1$ sedimentation is a significant effect, indeed we predict that particles should sediment $9\mu$m over the length of the channel.

The two algorithms produce consistent results in experiments where both algorithms can be used. The velocimetric algorithm has two major advantages over the laplacian algorithm.  First, the velocimetric algorithm is much more precise: we can measure heights to a precision of less than a micron, whereas the laplacian algorithm depends on the precision of the focal plane height (in this case, $3\mu$m).  Second, the velocimetric algorithm can be used at much larger Reynolds numbers than the laplacian algorithm since motion blur does not interfere with the height measurements.

\section*{Theory of inertial migration}

We adapt the asymptotic theory developed by Hood et al \cite{Hood15} for square channels to predict the inertial forces in rectangular channels. Since numerical experiments show that viscous stresses dominate momentum flux terms over the entire fluid filled domain, $V$, we can perform a regular perturbation expansion in the particle Reynolds number $\Rey_p$, treating the viscous and pressure stresses as dominant terms, and the inertial stress as a perturbative correction. 

We use the Lorentz reciprocal theorem \cite{Leal80} to represent the inertial lift force $\mathbf{F}_L$ as a volume integral that involves the following three solutions of Stokes equations ($\Rey_p = 0$): (1) $\bar{\bu}$, the undisturbed flow through the channel, (2) $\bu$, the solution for a force-free and torque-free sphere moving through the microchannel, and (3) a test velocity $\hat{\bu}$ for the slow ($\Rey_p = 0$) movement of a particle in the lateral direction in a quiescent fluid.  The total force on a particle that is constrained from migrating across streamlines can be written as an integral:
\begin{equation}
\mathbf{F}_L =  \Rey_p \int_V \hat{\bu} \cdot (\bar{\bu} \cdot \nabla \bu + \bu \cdot \nabla \bar{\bu} + \bu \cdot \nabla \bu ) \, \mathrm{dv}. \label{eq:reciprocalthm}
\end{equation}
To expose the role played by particle size in determining the lift force,we expanded $\bu$ and $\hat{\bu}$ as a two-term series in $\frac{a}{H}$, the ratio of the particle radius to the channel depth. The lift force $\mathbf{F}_L$ at the point $\bx_0$ in the channel can be expressed as a two term asymptotic expansion with coefficients $\mathbf{c}_4(\bx_0)$ and $\mathbf{c}_5(\bx_0)$.  Specifically,
\begin{equation}\label{eq:scaling_law}
      \mathbf{F}_L(\bx_0) \sim \frac{\rho U^2 a^4}{H^2} \left[ \mathbf{c}_4(\bx_0) + \frac{a}{H} \mathbf{c}_5(\bx_0)\right] ,
\end{equation}
where $\rho$ is the fluid density, $H$ is the channel depth, and $U$ is the average velocity of the undisturbed flow. The coefficients $\mathbf{c}_4(\bx_0)$ and $\mathbf{c}_5(\bx_0)$ are dimensionless constants including both analytical and numerically computed components, and that depend on the location of the particle $\bx_0$ and the aspect ratio of the rectangular cross-section.  A text file giving the values of $\mathbf{c}_4(\bx_0)$ and $\mathbf{c}_5(\bx_0)$ for a grid of particle locations is included in the supplemental materials.

The method above, which adapts the results from Hood et al \cite{Hood15} for a channel with aspect ratio two, gives only the focusing force on a particle that is not free to migrate across streamlines. The particles in our experiments are free to migrate under inertial focusing forces. We find the migration velocity $\bu_m = (u_m,v_m)$ of a force-free particle by equating the lift force (\ref{eq:scaling_law}) with the drag force computed for a particle translating with a general velocity $\bu_m$ \cite{HappelBrenner83}. This drag force can be evaluated by the method of reflections, to the same order of accuracy as equation (\ref{eq:scaling_law}):
\begin{equation}
       6\pi\mu a [\bu_m(\bx_0) + \bu_{im}(\bx_0)]= \mathbf{F}_L(\bx_0) ,
\end{equation}
where $\bu_{im}$ is the leading order backflow created at $\bx_0$ due to the walls of the microchannel. Furthermore, $\bu_{im}(\bx_0)$ is the first order correction calculated by the method of reflections for a small sphere migrating across streamlines and therefore is linearly related to the lift force $\mathbf{F}_L(\bx_0)$, namely there exists a matrix $\mathbf{S}(\bx_0)$ such that $\bu_{im}(\bx_0) \simeq \mathbf{S}(\bx_0)\cdot \mathbf{F}_L(\bx_0)$.  The terms of $\mathbf{S}(\bx_0)$ are determined by computing the reflection $\hat{\bu}_2$ of the test velocity $\hat{\bu}$ and evaluating at the center of the particle $\bx_0$.  More specifically, denote the method-of-reflections correction for a point force located at $\bx_0$ and and pointing in the direction $\mathbf{e}_i$ by $\hbu_{2,i}(\bx_0)$. In this case $\mathbf{S}(\bx_0) = S_{ij}(\bx_0) $ is defined as:
\begin{equation}\label{eq:trajectory}
     S_{ij}(\bx_0) = (\hbu_{2,i}(\bx_0) \cdot \mathbf{e}_{j}). 
\end{equation} 
Rearranging the terms above for the migration velocity gives:
\begin{equation} \label{eq:mobility}
      \bu_m(\bx_0) = \left[\mathbf{I} + \frac{a}{H} \mathbf{S}(\bx_0)\right] \frac{\mathbf{F}_L(\bx_0)}{6\pi\mu a}.
\end{equation}
The pre-factor here represents the tensorial mobility of the particle.

\begin{figure}
    \includegraphics[scale=.9]{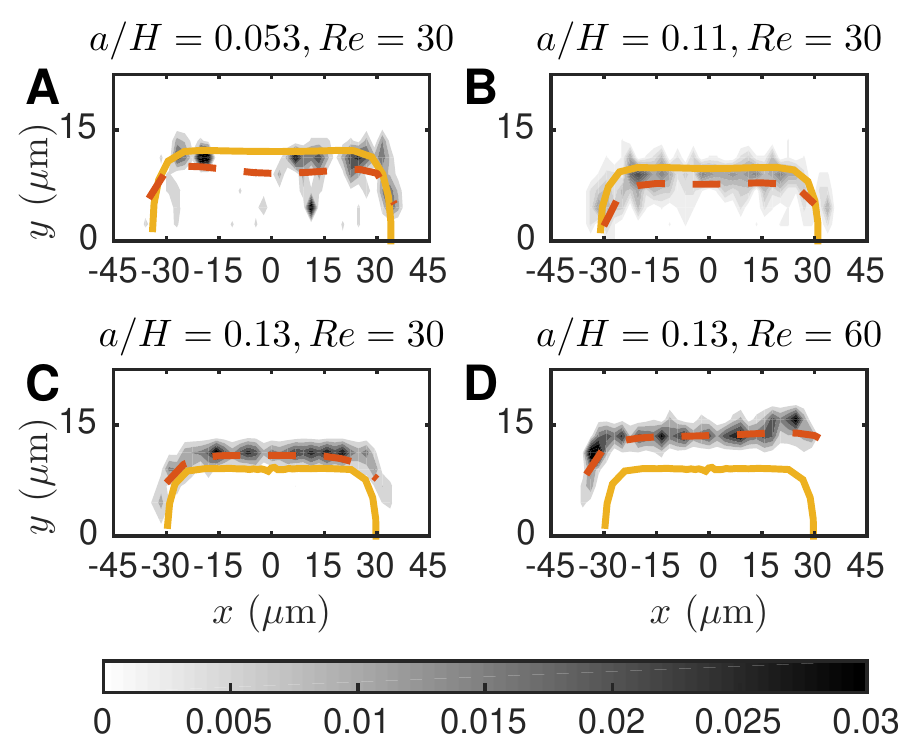}
   \caption{\label{fig:piv_manifold}  PDF of particle location in the upper half of the channel, with the gray scale indicating density. The predicted manifold (solid yellow line) is a good approximation of the measured manifold (dashed orange line). The particle size and Reynolds number in each figure are: (A) $a/H = 0.053$ $\Rey = 30$; (B) $a/H = 0.11$ $\Rey = 30$; (C) $a/H = 0.13$ $\Rey = 30$; and (D) $a/H = 0.13$ $\Rey = 60$. }
\end{figure}

We are interested in how particles travel due to this migration velocity, which can be computed at any point $\bx_0$ in the channel. Let $\mathbf{X}(t)=(X(t),Y(t))$ be the location of a given particle in the channel cross-section as a function of time $t$.  For a particle migrating due to inertial lift forces:
\begin{equation}\label{eq:advect}
      \frac{d\mathbf{X}}{dt} = \bu_m \, , \qquad \mathbf{X}(0) = (x_0,y_0).
\end{equation}
The slow-focusing manifold is evaluated numerically by advecting particles according to (\ref{eq:advect}) and finding the curve $\Lambda$ which is invariant under (\ref{eq:advect}).  Note that $\Lambda$ depends on the relative particle size $\frac{a}{H}$.  At any point $\bx_0$ in the channel, the migration velocity satisfies
\begin{equation}\label{eq:migvelgen}
     \mathbf{u}_m(\bx_0) \sim \frac{\rho U^2 a^3}{6 \pi \mu H^2}  \left[ \mathbf{I} + \frac{a}{H} \mathbf{S}(\bx_0) \right] \cdot \left[ \mathbf{c}_4(\bx_0) + \frac{a}{H} \mathbf{c}_5(\bx_0)\right].
\end{equation}
where the coefficients $\mathbf{c}_4(\bx_0)$ and $\mathbf{c}_5(\bx_0)$ are the same as those calculated in (\ref{eq:scaling_law}). 

The limiting assumptions in the development of equation (\ref{eq:migvelgen}) are twofold: (i) in order to make our regular perturbation expansion we assume $\Rey_p \ll 1$ and (ii) in order to represent the particle by a singularity we assume that the particle is much smaller than $h$ the distance from the particle to the wall, $a \ll h \sim \frac{1}{6}H$.  However, in practice conditions (i) and (ii) can be relaxed to a larger set of values for $\Rey_p$ and $\alpha$. Hood et al\cite{Hood15} show that, because the presence of the walls diminishes the size of the inertial term in the NSE, empirically this model is accurate up to $\Rey_p \le 7$.  Furthermore, Hood et al\cite{Hood15} empirically that the particle size limitation can be relaxed to $\alpha \le 0.2$.  In our experiments we have $\Rey_p \le 3.2$ and $\alpha \le 0.21$, so equation (\ref{eq:migvelgen}) should be a good approximation of the migration velocity.

\begin{figure}[t]
    \includegraphics[scale=1]{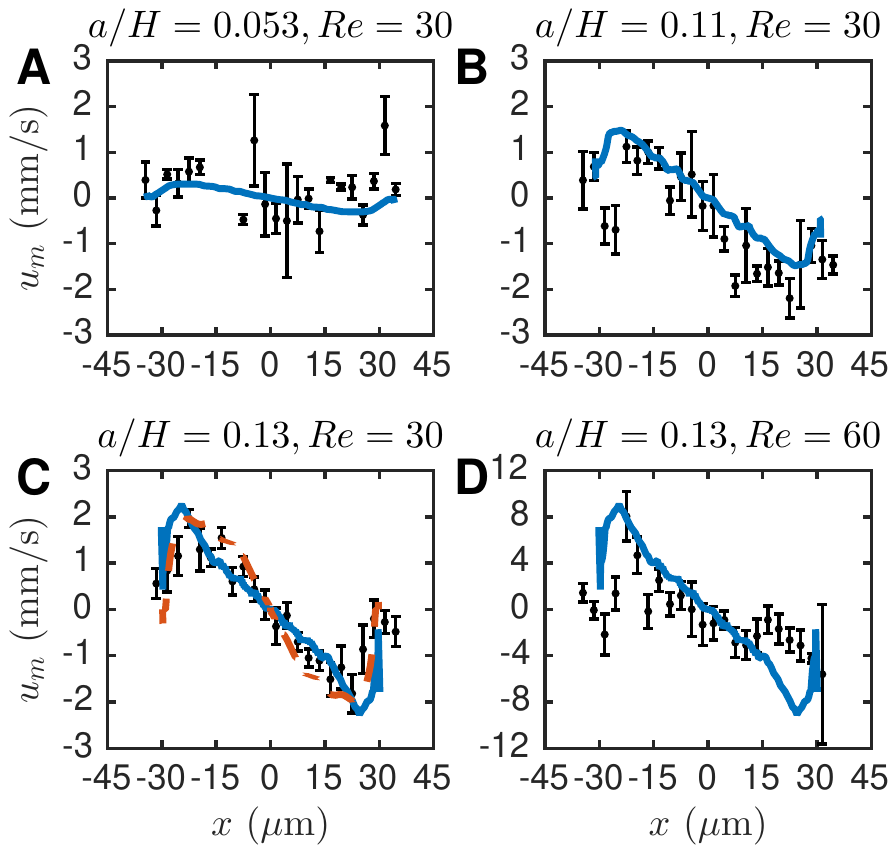}
   \caption{\label{fig:piv_vel} The measured migration velocity along the measured manifold (black markers) agrees quantitatively with the asymptotic theory (blue line) in equation (\ref{eq:migvelgen}) and numerical solution of the NSE (orange dashed line).  The particle size and Reynolds number in each figure are: (A) $a/H = 0.053$ $\Rey = 30$; (B) $a/H = 0.11$ $\Rey = 30$; (C) $a/H = 0.13$ $\Rey = 30$; and (D) $a/H = 0.13$ $\Rey = 60$.}
\end{figure} 

The prediction of the focusing manifold $\Lambda$ compares well to the measured manifold in experiments (Fig. \ref{fig:piv_manifold}A-D).  The measured manifold is found by fitting a quadratic polynomial to the measured $(x,y)$ locations of all the particles.  Even though our theory assumes that $\Rey_p \ll 1$, the predicted manifold $\Lambda$ is a fair approximation even when $\Rey_p = 1.01$ (Fig. \ref{fig:piv_manifold}D).  Additionally, deformation of the PDMS channel has been reported at higher Reynolds numbers \cite{sollier2011rapid}, which is not taken into account in our theory.

Lateral migration velocities along the manifold quantitatively agree with the asymptotic theory in equation (\ref{eq:migvelgen}). We filtered the measured velocities to select particles that were within a distance $2.25 \mu$m of the slow manifold. We then binned these particles into $3\,\mu$m $x-$intervals, and averaged migration velocities for particles within the same bin. Experimental measurements of migration velocity along the slow manifold agree almost exactly with the asymptotic prediction of the migration velocity along the theoretical manifold (Fig. \ref{fig:piv_vel}A-D) including different particle sizes and flow speeds.

There are no free parameters in the prediction of the migration velocity in equation (\ref{eq:migvelgen}). The asymptotic result supports that $\bu_m\propto U^2$, just as was found in previous numerical simulations \cite{DiCarlo09}. The asymptotic theory also shows that migration velocity has no clear power law dependence on particle size. This asymptotic theory is most accurate for small particle sizes and moderate Reynolds numbers; in practice requiring that $\frac{a}{H}<0.2$, and that channel Reynolds number $\Rey \lesssim 80$.

\section*{Dependence of focusing forces on particle size and Reynolds number}

We performed similar analysis of migration velocities for particles of different sizes and for different flow velocities.  Note that the migration velocity is a vector field $\bu_m = (u_m,v_m)$, and recall that in our experimental setup, we can only measure the slow phase of inertial migration.  This corresponds to measuring the $x$-component $u_m$ of the migration along the manifold. We define the average migration velocity $\langle u_m \rangle$ as the average of $-\sign(x)u_m$ over all bins, where $u_m$ is first averaged in each bin. The $-\sign(x)$ factor prevents left and right sides of the channel from canceling since $u_m$ is an odd function across $x=0$.

\begin{figure}
    \includegraphics[scale=1]{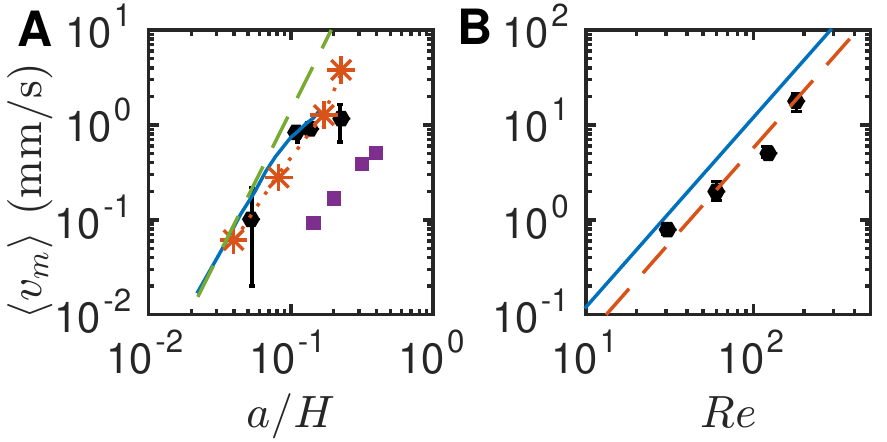}
   \caption{\label{fig:alpha} (A) Over the range of measured particle sizes there is no simple power law for the dependence of migration velocity upon particle size, $a$. Here we fixed $Re = 30$ and varied particle diameter (dashed green line: $a^3$ scaling law, blue line: Equation (\ref{eq:migvelgen}), black circles: measured average migration velocity $\pm$ s.e., orange stars: numerical prediction of average migration velocity).  Zhou and Papautsky's \cite{Papautsky13} indirect measurements (purple squares) show a similar trend, but are an order of magnitude smaller. (B) Average migration velocities scale like $U^2$. Here we fixed particle diameter at $d = 12 \mu m$ and varied the flow rate (blue line: Equation (\ref{eq:migvelgen}), dashed orange line: numerical fit of $U^2$ with one free parameter, black circles - measured average migration velocity$\pm$ s.e.).}
\end{figure}

Average migration velocity $\langle u_m \rangle$ does not have a power law dependence upon particle size $a$, but agrees quantitatively with (\ref{eq:migvelgen}). For very small particles, migration velocities increase with $a^3$ scaling law, as predicted asymptotically \cite{HoLeal74,Hinch89}, but this power law breaks down even at small particle sizes. Incorporating an extra term in the series expansion produces good fit up to $\frac{a}{H}= 0.16$ in our data. To clarify that there is no conflict between numerical data and experimental data we computed the migration forces on a particle using the same finite element simulation that was used to extract the downstream velocity of the particle over a range of particle sizes ($\frac{a}{H} = 0.04, 0.08, 0.17,$ and $0.23$) that covered the entire experimental range. Numerical migration velocities averaged over the slow manifold agreed with experimental measurements and, over their range of validity, with the asymptotic series also (Fig. \ref{fig:alpha}A).

Migration velocities scale like $U^2$. Asymptotic studies agree \cite{HoLeal74, Hinch89, Hood15} that if particle size is fixed while the flow rate through the microchannel is varied then since in (\ref{eq:reciprocalthm}) both $\bu$ and $\bar{\bu}$ vary in proportion to $U$, the total migration force $\mathbf{F}_L$ and total migration velocity $u_m$ will scale like $U^2$. Our experimental measurements confirm this scaling (Fig. \ref{fig:alpha}B). Experiments at much higher Reynolds numbers have shown that additional focusing positions appear in channel corners \cite{Gijs2013,Sugihara-Seki2014}, but we find no evidence of alternate focusing positions over the range $\Rey=30-180$.  

\begin{figure}
    \includegraphics[scale=.95]{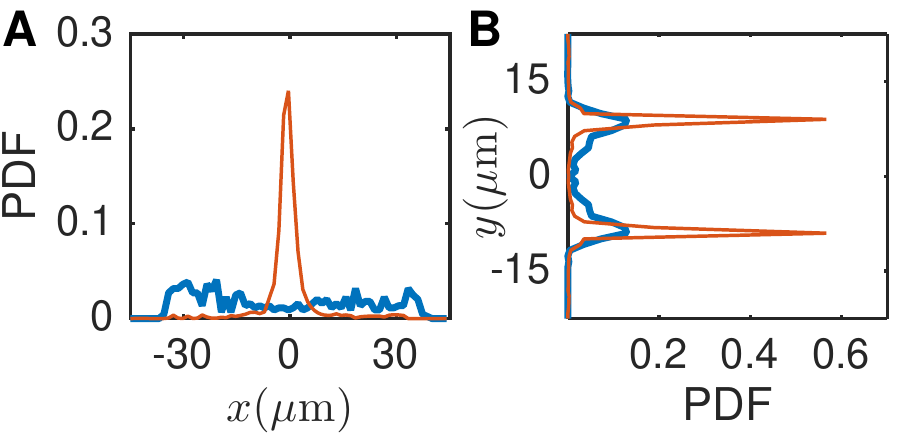}
   \caption{\label{fig:inlet} Particles enter the microchannel prefocused to a thin band of $y-$ coordinates, so only slow focusing dynamics can be measured. (A) The particle $x$-position PDF is nearly uniform at channel entry (thick blue line) becoming focused after traveling $1.5$cm through the channel (orange line). (B) However, the particle $y$-position PDF is strongly focused both at entry (blue), and after particles have reached their focusing streamline.  Recall that the reconstruction algorithm cannot decipher between $+y$ and $-y$ values, we have made the distribution symmetric to illustrate that both positive and negative $y-$values can be achieved. (Relative particle size $a/H=0.11$, channel Reynolds number $\Rey=30$). \label{fig:hist}}
   %The critical inlet length $L_i$ (blue line) is and the development length $L_d$ (dashed green line) intersect at $W_i/W_0 = 30$, for $12 \mu m$ particles and $\Rey = 30$.}
\end{figure}

Our direct measurements of particle migration show that asymptotic theory adapted for rectangular micro-channels can quantitatively predict inertial lift forces on particles, including their dependence on particle size and channel velocity. Why have  indirect measurements of migration velocities by Zhou and Papautsky \cite{Papautsky13} contradicted theory? First we note that our inertial migrational velocities are an order of magnitude larger than previous experiments (Fig. \ref{fig:alpha}A), likely because indirect focusing measurements do not equally weight trajectories across the entire slow manifold, but rather only the slowest focusing that occurs as particles approach the focusing streamline. Additionally, Zhou and Papautsky \cite{Papautsky13} assume that particles are uniformly spread across the microchannel cross-section before focusing. We found that particles appeared to be uniformly dispersed (Fig. \ref{fig:hist}A) at the inlet. However, our reconstruction of particle depth showed that particles entered the microchannel already focused in their $y$-coordinate (Fig. \ref{fig:channel}B and \ref{fig:hist}B). Thus, our in-channel measurements showed only the second phase of inertial migration along a single slow manifold. Thus, pre-focusing makes it impossible to separate fast and slow phases of focusing in the manner attempted by Zhou and Papautsky \cite{Papautsky13}.

\section*{Pre-focusing in the channel inlet}

Pre-focusing is due to inertial lift forces acting in the channel inlet. We can use asymptotic theory to predict the ammount of prefocusing, which occurs primarily in the depth ($y$-) dimension where velocity shear is largest. In this section we will derive an expression for the $y$-distance a particle migrates in the channel inlet.

We model the inlet region as a linear contraction in the $x-$direction, with maximum width $W_i$ at $z=-L_i$ and minimum width $W_0$ at the opening of the channel at $z=0$, and constant depth $H$ (Fig. \ref{fig:inlet}). Assuming constant flow rate $Q$ throughout the channel, and self-similar velocity profiles across each cross-section of the channel inlet, the downstream characteristic velocity in the inlet region takes the form: $U(z) = \frac{U_0 W_0}{W(z)}$, where $W(z)$ is the width of the channel inlet, specifically,
\begin{equation}
      W(z) = W_0 - \frac{z}{L_i}(W_i - W_0).
\end{equation}
For a particle lying on the symmetry plane $x=0$, then the time-evolution of the $y$-component of the particle location obeys the ODE:
\begin{equation}
      \frac{dy}{dt} = v_m(x=0,y) \sim \frac{\rho U^2 a^3}{6 \pi \mu H^2} c_L(x=0,y).
\end{equation}
Here we take the first order approximation of the migration velocity $\bu_m = (u_m,v_m)$ in equation (\ref{eq:migvelgen}).  By Taylor expanding the migration velocity around the equilibrium position $y_{eq}$, and making the change of variables $Y = y-y_{eq}$ we obtain the following ODE:
\begin{equation}\label{eq:ode1}
      \dot{Y} = -\Gamma(z) Y,
\end{equation}
where $-\Gamma(z) = \frac{d}{dy} v_m$. Let $\Gamma_0 = \Gamma(0)$ be the rate of change of the migration velocity at the widest point of the channel $z=0$, then since the migration velocity scales with $U^2$ we have:
\begin{equation}
      \Gamma(z) = \frac{W_0^2}{W(z)^2}\Gamma_0.
\end{equation}
So:
\begin{equation}
      \frac{dY}{dz} \frac{dz}{dt} = \frac{dY}{dz} \frac{U_0 W_0}{W(z)} = -\frac{W_0^2}{W(z)^2}\Gamma_0 Y
\end{equation}
Integrating and rearranging gives:
\begin{equation}\label{eq:prefocusing}
      \frac{Y_0}{Y_i} = \left(\frac{W_0}{W_i}\right)^{\eta L_i} \quad \mbox{, where} \quad \eta = \frac{\Gamma_0 W_0}{U_0(W_i-W_0)}.
\end{equation}
From equation (\ref{eq:migvelgen}) we estimate:
\begin{equation}
      \Gamma_0 = -120.3 \left( \frac{a^3 \Rey U_0}{6 \pi H^4}\right).
\end{equation}

Using the channel dimensions from this experiment, with $Re = 30$ and $a = 5 \mu$m, we find that particles are within $1.5 \mu$m of the equilibrium position $y_{eq}$ by the end of the inlet region, $z=0$, consistent with our measurements (Fig. \ref{fig:hist}B). However, little focusing occurs in the $x-$direction, so that if particle $x-$ positions only are measured, as in Zhou \& Papautsky \cite{Papautsky13} particles appear to be uniformly dispersed across the channel (Fig. \ref{fig:hist}A).

\begin{figure}
   \centerline{
          \includegraphics[scale=.4]{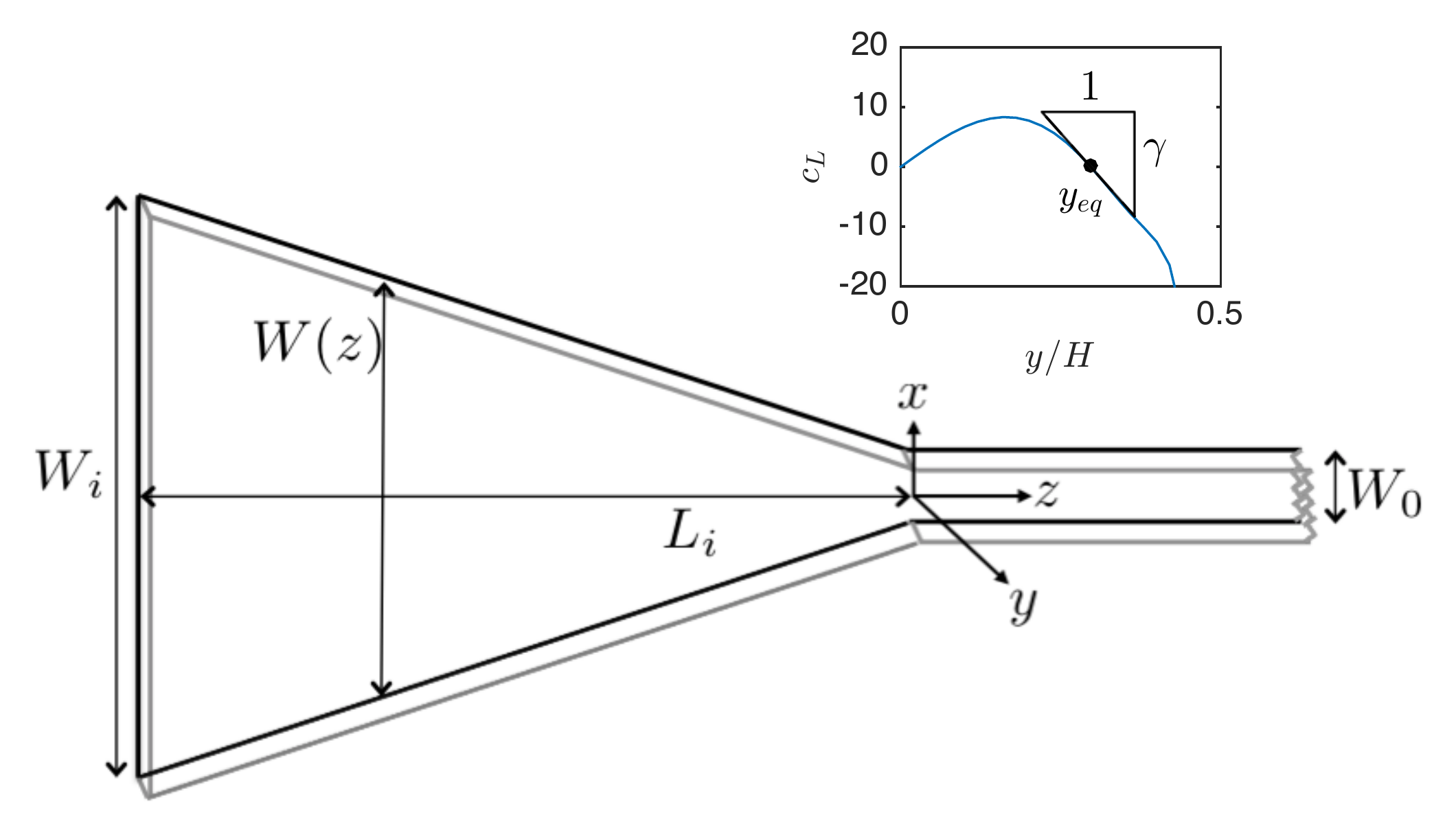}
%          $(b)$\includegraphics[scale=1]{gamma0_plot.eps}
   }
   \caption{\label{fig:inlet}  Diagram of inlet region (not to scale). (Inset) Plot of $c_L$, the particle lift force coefficient, the slope of the tangent line at the equilibrium focusing depth is $\gamma = -120.3$.}
\end{figure}

Can a microchannel inlet be designed to measure fast-focusing dynamics? Equation (\ref{eq:prefocusing}) shows that shorter inlet regions (smaller values of $L_i$) lead to less particle pre-focusing. To enforce that focusing produces a less than 10\% disturbance of particle depths during their passage through the inlet, i.e. that $\frac{Y_0}{Y_i}>0.9$, we invert (\ref{eq:prefocusing}) and find that if the particle radius $a$ is measured in microns, then the maximum inlet length, also in microns, is given by $L_i=2100/a^3$. In particular for a particle with radius $a = 5\mu$m, the maximum channel inlet length is only $L_i=17\,\mu$m. 

However, to see fast-focusing dynamics there must also be fully developed Poiseuille flow at the channel inlet. The inlet must therefore be longer than the development length, $L_d$, required for viscous boundary layers to diffuse from the channel floor and ceiling and to fill the entire channel. Ciftlik et al\cite{Gijs2013} give $L_d = \frac{1}{30} \Rey H = 45\,\mu$m at the lowest Reynolds numbers used in our experiments, exceeding the minimum $L_i$. These competing constraints make it impossible to design a microchannel inlet to measure fast focusing dynamics. Fast focusing dynamics can nevertheless be observed in glass capillaries \cite{ChoiSeoLee11} where inlet regions can be removed, however glass microfluidic capillaries can not be machined into {\it de novo} geometries.

\section*{Conclusions}

The first reported experimental measurements of inertial migration velocities show that there is no conflict between asymptotic theory and the measured inertial migration velocities of particles in microchannels. However, a theory capable of quantitatively describing these forces does not produce a simple power law dependence of migration velocities upon particle size, contributing to previous contradictions between experiments, numerical data and theory. Additionally, we show that in soft lithography microchannels, fast focusing dynamics occur in the channel inlet, causing pre-focusing of particles before they enter the microchannel imposing previously unexamined constraints over the control that can be exerted over particle focusing trajectories.

\section*{Acknowledgements}
This work was partly supported by the National Science Foundation through grants DGE-1144087 (to K. Hood) and DMS-1312543 (to M. Roper).  This work was partly supported by the Office of Naval Research Young Investigator Program Grant \# N000141210847 (to D. Di Carlo).  This work was partially supported by the UCLA Dissertation Year Fellowship (to K. Hood).

\section*{Appendix: Asymptotic Calculation of Migration Velocity}\label{appendix}

Here we provide more detail for the asymptotic calculation of the inertial lift force (Equation (\ref{eq:scaling_law}) in the main text).  The details of this calculation are already described in Hood et al \cite{Hood15} so our treatment here emphasizes the modifications nedded for a channel with non-square cross-section.  We consider a single spherical particle  of radius $a$ suspended in a rectangular channel with aspect ratio two.  The origin is located at the center of the particle, and the particle is allowed to translate downstream with velocity $\mathbf{U}_p = U_p \ez$ and to rotate with angular velocity ${\Omega}_p$.  $\mathbf{U}_p$ and ${\Omega}_p$ are chosen so that the particle is totally torque free and force free in the downstream direction. It will in general experience forces in the $x-$ and $y-$ direction. From these forces we can compute the migration velocity for a particle that is totally force and torque free.

First we define the three-dimensional undisturbed flow, $\mathbf{\bar{u}}$, which is rectangular channel Poiseuille flow\citep{Papanastasiou99} with centerline velocity $U$, width $W$, and height $H$, and takes the form $\mathbf{\bar{u}}= \bar{u}(x,y)\mathbf{e}_{z}$, where $\mathbf{e}_z$ is a unit vector pointing in the downstream direction. The velocity $\mathbf{\bar{u}}$ and pressure $\bar{p}$ solve the Stokes equations with boundary condition $\mathbf{\bar{u}} = \mathbf{0}$ on the channel walls.  We will also need the Taylor series expansion for $\bar{u}$ around the center of the particle:
\begin{align}\label{eq:ubarseries}
      \bar{u}(x,y) = \beta + \gamma_x x + \gamma_y y + \delta_{xx} x^2 + \delta_{xy} xy +\delta_{yy} y^2  \\ + O(x^3,y^3,xy^2,x^2y) \nonumber
\end{align}
where we define our origin of coordinates to coincide with the center of the sphere.

Within the microchannel, the fluid velocity $\bu$ and pressure $p$ are governed by the dimensionless steady-state 3D Navier-Stokes Equations (NSE) in the reference frame of the moving particle:
\begin{eqnarray}\label{eq:NSE}
      \nabla^2 \bu - \nabla p &=& \alpha^2 \Rey (\bar{\bu} \cdot \nabla \bu + \bu \cdot \nabla \bar{\bu} + \bu \cdot \nabla \bu) \, , \nonumber \\
      \nabla \cdot \bu &=& 0 \, , \\
      \bu &=& \mathbf{U}_p + {\Omega}_p \times \mathbf{r} - \bar{\bu} \quad \textrm{on} \quad |\mathbf{r}| = 1 \, , \nonumber \\
      \bu &=& 0 \quad \textrm{on the channel walls, and as } z\to \pm \infty \,  .\nonumber
\end{eqnarray}
The dimensionless equations are obtained by scaling lengths by the particle radius $a$, velocities by the velocity $U a/H$, and pressures are scaled by $\mu U/H$ where $\mu$ is the dynamic viscosity. 

The calculation of the inertial lift force $\mathbf{F}_L$, and consequently the migration velocity $\bu_m$ is outlined as follows.  First we make a regular perturbation expansion in the particle Reynolds number $\Rey_p$ and use the Lorentz reciprocal theorem to represent the lift force $\mathbf{F}_L$ in terms of the perturbation expansion.  Then we further expand the terms in the reciprocal theorem integral as a series expansion in the relative particle size $\alpha = \frac{a}{H}$, assumed to be asymptotically small. As $\alpha\to 0$, the reciprocal theorem integral must be calculated by dividing it into two subdomains, in which different terms dominate within the integrand, we call the contributions from these two regions the inner and outer integrals. We must combine the inner and outer integrals to find the inertial lift force $\mathbf{F}_L$.  

\subsection*{Perturbation Expansion}

For small particle sizes, the particle Reynolds number $\Rey_p = \alpha^2 \Rey$ is a small parameter.  While {\it a priori} estimates suggest that inertial stresses will become co-dominant with viscous and pressure forces sufficiently far from the particle \cite{Hinch89}, numerical examination of the terms of (\ref{eq:NSE})  shows that the inertia is subdominant throughout the channel; because of this, we can treat inertial stresses as a small perturbation to the solution produced by balancing viscous and pressure stresses across the entire channel cross-section, i.e. perform a regular perturbation expansion in $\Rey_p$ \cite{Hood15}.  We then expand further in the small parameter $\alpha$, following for this second part, the method proposed by Ho \& Leal \cite{HoLeal74}, but using numerical PDE methods to compute boundary corrections that arise in the solution, and extending the solution to include the next correction from $\alpha$, to capture the fact that the particle migration velocity has no simple power law dependence on particle size.

We expand the fluid velocity $\bu$, pressure $p$, particle velocity $\mathbf{U}_p$, and particle rotation ${\Omega}_p$ in the small parameter $\Rey_p$,
\begin{align}
      \bu = \bu^{(0)} + \Rey_p \bu^{(1)}  + \hdots\, , \\
      \quad  p = p^{(0)} + \Rey_p p^{(1)} + \hdots \, , \quad \mathrm{ etc.,} \nonumber
%      \mathbf{U}_p &=& \mathbf{U}_p^{(0)} + \Rey_p \mathbf{U}_p^{(1)} + \Rey_p^2 \mathbf{U}_p^{(2)} + \hdots \, , \mathrm{ and} \\ 
%      \mathbf{\Omega}_p &=& \mathbf{\Omega}_p^{(0)} + \Rey_p \mathbf{\Omega}_p^{(1)} + \Rey_p^2 \mathbf{\Omega}_p^{(2)} \hdots \, , \nonumber
\end{align}
and substitute into (\ref{eq:NSE}) and collect like terms in $\Rey_p$.  The first order velocity and pressure solve the homogeneous Stokes problem:
%\begin{equation}
  \begin{align}\label{eq:first_order}
  \nabla^2 \mathbf{u}^{(0)} - \nabla p^{(0)} &= \mathbf{0} , \qquad 
  \nabla \cdot \mathbf{u}^{(0)} = 0, \nonumber \\ 
  \mathbf{u}^{(0)} &= \mathbf{U_p}^{(0)} + {\Omega_p}^{(0)} \times \mathbf{r} - \mathbf{\bar{u}}\mbox{ on }  r = 1, \\ 
   \mathbf{u}^{(0)} &= \mathbf{0} \mbox{ on channel walls and as } z \to \pm \infty, \nonumber
   \end{align}
%\end{equation}
while the second order velocity and pressure solve the inhomogeneous Stokes problem:
\begin{eqnarray}\label{eq:second_order}
  \nabla^2 \mathbf{u}^{(1)} - \nabla p^{(1)} &=& ( \mathbf{\bar{u}} \cdot \nabla \mathbf{u}^{(0)} + \mathbf{u}^{(0)} \cdot \nabla \mathbf{\bar{u}}  + \mathbf{u}^{(0)} \cdot \nabla \mathbf{u}^{(0)} ) , \nonumber \\ 
  \nabla \cdot \mathbf{u}^{(1)} &=& 0, \nonumber \\
  \mathbf{u}^{(1)} &=& \mathbf{U_p}^{(1)} + {\Omega_p}^{(1)} \times \mathbf{r} \mbox{ on }  r = 1, \label{eq:u1} \\
   \mathbf{u}^{(1)} &=& \mathbf{0} \mbox{ on channel walls and as } z \to \pm \infty. \nonumber
\end{eqnarray}
This is a regular perturbation expansion: the right hand side of (\ref{eq:second_order}) is the inertial stress associated with the solution of (\ref{eq:first_order}).

Since only the force on the particle is required, and not the complete velocity field $\mathbf{u}^{(1)}$, we can use the Lorentz Reciprocal Theorem \cite{Leal80}, to express the inertial lift force $\mathbf{F}_L$ as an integral containing only a solution of (\ref{eq:first_order}) $\bu^{(0)}$:
\begin{align}\label{eq:recipthm}
      \mathbf{e} \cdot \mathbf{F}_L =  \int_V & \mathbf{\hat{u}}  \cdot \\  &\left(  \mathbf{\bar{u}} \cdot \nabla \mathbf{u}^{(0)} + \mathbf{u}^{(0)} \cdot \nabla \mathbf{\bar{u}}  + \mathbf{u}^{(0)} \cdot \nabla \mathbf{u}^{(0)}  \right)  \mathrm{dv}. \nonumber
\end{align}
Here to calculate the lift force acting on the particle in the direction $\mathbf{e}$ we must integrate the inertial stresses against the Stokes ($\Rey=0$) solution, $\hat{\bu}$, for the same particle moving at unit velocity in the the direction $\mathbf{e}$ in a quiescent fluid.  In other words $\hat{\bu}$ and an associated pressure $\hat{p}$ solve the homogenous Stokes problem:
\begin{eqnarray}
      \nabla^2 \hat{\bu} - \nabla \hat{p} &=& \mathbf{0} , \qquad 
  \nabla \cdot \hat{\bu} = 0, \nonumber \\
  \hat{\bu} &=& \mathbf{e} \mbox{ on }  r = 1, \\
  \hat{\bu} &=& \mathbf{0} \mbox{ on channel walls and as } z \to \pm \infty. \nonumber
\end{eqnarray}
If the particle size is known this method reduces the complexity of finding the focusing force from solving a nonlinear Navier-Stokes problem for $\bu$ to solving two linear homogenous Stokes problems for $\bu^{(0)}$ and $\hat{\bu}$.  However, the dependence of force upon particle size is not made explicit in the solution, and we analyze the equations in the limit where $\alpha \ll 1$ to find this dependence.

\subsection*{Series Expansion in $\alpha$}

We expand the velocities $\bu^{(0)}$ and $\hat{\bu}$ as power series in $\alpha$ using the method of reflections. 

Specifically, we follow Ho \& Leal\cite{HoLeal74} and Happel \& Brenner \cite{HappelBrenner83} and expand each velocity field as a sum of corrections:
\begin{eqnarray}
  \mathbf{u}^{(0)} &=& \mathbf{u}^{(0)}_1 + \mathbf{u}^{(0)}_2 + \mathbf{u}^{(0)}_3 + \mathbf{u}^{(0)}_4 + \ldots \, ,
\end{eqnarray}
with similar expansions for $p$, $\hbu$, and $\hp$.  Here, $\bu^{(0)}_1$ is the Stokes solution for a particle in unbounded flow (ignoring the channel walls),  $\mathbf{u}^{(0)}_2$ is the Stokes solution with boundary condition $  \mathbf{u}^{(0)}_2 = -\mathbf{u}^{(0)}_1$ applied on the channel walls (but ignoring the particle boundaries), and $\bu^{(0)}_3$ is the unbounded Stokes solution with boundary condition $\mathbf{u}^{(0)}_3 = -\mathbf{u}^{(0)}_2$ on the particle surface, etc.  Odd terms impose the boundary conditions on the particle, whereas even terms impose the boundary conditions on the channel walls.  %We will show below that the terms in this series constitute a power series in $\alpha$.

The first term in the series, $\bu^{(0)}_1$, is the solution for a particle in unbounded flow, can be found analytically using the Lamb's solution \cite{Lamb45,KimKarrila2005}.  Note that we have corrected an error from Hood et al \cite{Hood15} in the series below:
\begin{align*}
%      \begin{split}
              \bu = &- \frac{5\alpha z \br}{2r^2}\left(\frac{x}{r}\gx + \frac{y}{r}\gy 
                          \right)\frac{1}{r^2} \\
                    &+ \frac{\alpha^2 \dxx}{8} \left( \frac{5}{3}\ez -3\frac{x^2}{r^2}\ez       
                          +10 \frac{xz}{r^2}\ex +5\frac{z\br}{r^2} -35\frac{x^2z\br}{r^4} 
                          \right)\frac{1}{r^3} \\
                    &+ \frac{\alpha^2\dxy}{8}\left( -3\frac{xy}{r^2}\ez 
                          + 5\frac{yz}{r^2}\ex + 5\frac{xz}{r^2}\ey 
                          - 35\frac{xyz\br}{r^4} \right)\frac{1}{r^3} \\
                    &+ \frac{\alpha^2\dyy}{8} \left(\frac{5}{3}\ez - 3\frac{y^2}{r^2}\ez 
                          +10\frac{yz}{r^2}\ey 
                          -35\frac{y^2z\br}{r^4}  \right)\frac{1}{r^3} \tag{\stepcounter{equation}\theequation} \\
                    &- \frac{\alpha\gx}{2}\left( \frac{z}{r}\ex + \frac{x}{r}\ez 
                          - 5\frac{xz\br}{r^3} \right)\frac{1}{r^4} \\
                    &- \frac{\alpha\gy}{2}\left( \frac{z}{r}\ey + \frac{y}{r}\ez 
                          - 5\frac{yz\br}{r^3} \right)\frac{1}{r^4} \\
                    &+ \frac{\alpha^2\dxx}{8}\left( \ez - 5\frac{x^2}{r^2}\ez 
                          - 10\frac{xz}{r^2}\ex - 5\frac{z\br}{r^2} 
                          +35\frac{x^2z\br}{r^4} \right) \frac{1}{r^5} \\
                    &+\frac{\alpha^2\dxy}{8}\left( -5\frac{yz}{r^2}\ex 
                          - 5\frac{xz}{r^2}\ey - 5\frac{xy}{r^2}\ez 
                          + 35\frac{xyz\br}{r^4} \right)\frac{1}{r^5} \\
                    &+\frac{\alpha^2\dyy}{8}\left( \ez - 5\frac{y^2}{r^2}\ez 
                          - 10\frac{yz}{r^2}\ey - 5\frac{z\br}{r^2} 
                          + 35\frac{y^2z\br}{r^4} \right)\frac{1}{r^5}.
%      \end{split}
\end{align*}
Likewise, $\hbu_1$ can be calculated explicitly. Assuming that $\mathbf{e} = \ey$, then:
\begin{equation}
      \hbu_1 = \frac{3}{4}\left(\mathbf{e}_y+\frac{y\mathbf{r}}{r^2}\right) \frac{1}{r} + \frac{1}{4}\left(\mathbf{e}_y - \frac{3y\mathbf{r}}{r^2}\right) \frac{1}{r^3}.
\end{equation}
The remaining odd order terms can be found similarly. The even terms in the series expansions of $\bu^{(0)}$ and $\hbu$ are found numerically using a Finite Element Model implemented in Comsol Multiphysics (Comsol, Los Angeles).

\subsection*{Evaluation of the reciprocal theorem integral}

Given the Stokes velocities $\bu^{(0)}$ and $\hbu$ we can compute the inertial lift force $\mathbf{F}_L$ up to terms of $O(\Rey_p)$ using the reciprocal theorem (\ref{eq:recipthm}). It is advantageous to divide the fluid filled domain $V$ into two subdomains, $V_1$ and $V_2$, where:
\begin{equation}
  V_1 = \{ \mathbf{r} \in  V : r \le \xi \} \,\,\, \hbox{and} \,\,\, V_2 = \{ \mathbf{r} \in V : r \ge \xi\}.
\end{equation}
The intermediate radius $\xi$ is any parameter satisfying $1 \ll \xi \ll \frac{1}{\alpha}$.  Call the corresponding integrals the inner integral and the outer integral, and identify their contributions to the lift force as $\mathbf{F}_{L_1}$ and $\mathbf{F}_{L_2}$, respectively ($\mathbf{F}_L = \mathbf{F}_{L_1} + \mathbf{F}_{L_2}$).  The division of the integral into inner and outer regions allows one to incorporate varying length scales ($a$ for the inner region and $\ell$ for the outer region) into our model. Note that, distinct from Schonberg \& Hinch \cite{Hinch89}, inertia remains subdominant even in the outer region $V_2$. We will separately consider the contributions from the inner and outer integrals.

\subsection*{Inner Integral}
Since the odd terms in the method-of-reflections expansions for $\bu^{(0)}$ and $\hbu$ are prescribed on the boundary of the particle, each gives rise to several terms that contribute to the inner integral $\mathbf{F}_{L_1}$. By contrast, the outer terms influence $\mathbf{U}_p$ and ${\Omega}_p$, but do not contribute to the inner integrals directly. Since the odd terms are derived analytically from the Lamb's solution, it follows that $\mathbf{F}_{L_1}$ can also be computed analytically. We continue to scale lengths by $a$, so that $1\le r \le \xi\ll \alpha^{-1}$.  The inner integral can be expressed as the following expansion in $\alpha$.:
\begin{equation}
  \mathbf{F}_{L_1} = \rho U^2 a^2 ( \mathbf{h}_4 \alpha^2 + \mathbf{h}_5 \alpha^3 + \hdots \, )~~.
\end{equation}

In order to calculate the terms $\mathbf{h}_4$ and $\mathbf{h}_5$, we sort the terms of the Stokes velocities by leading order in $\alpha$.  We refer the interested reader to the authors' previous work \cite{Hood15} for the details of this calculation.  The first order contribution evaluates to zero, $\mathbf{h}_4 = \mathbf{0}$.  The next order contribution $\mathbf{h}_5 = (h_{5,x}, h_{5,y})$ is listed below (note that we correct an error from \cite{Hood15}):
\begin{align}
      h_{5,x} &= \frac{26 \pi \gamma_x \delta_{xx}}{9} + \frac{11 \pi \gamma_y \delta_{xy}}{12} + \frac{19 \pi \gamma_x \delta_{yy}}{18}, \\
      h_{5,y} &=  \frac{26 \pi \gamma_y \delta_{yy}}{9} + \frac{11 \pi \gamma_x \delta_{xy}}{12} + \frac{19 \pi \gamma_y \delta_{xx}}{18}.
\end{align}

\subsection*{Outer Integral}
For the outer integral we will consider alternate dimensionless variables, by using the rescaled distance $\mathbf{R} = \alpha \mathbf{r}$. This corresponds to using $H$ to non-dimensionalize lengths, rather than $a$.  We call these variables the outer variables, and we will denote them with uppercase roman letters.  In the outer region $V_2$, we must express our functions in terms of $\mathbf{R}$ and rearrange our functions by order of magnitude in $\alpha$.  Then the reciprocal theorem integral takes the following dimensional form:
\begin{align}
  f_{L_2}= \rho U_m^2 & \ell^2  \int_{V_C}  \mathbf{\hat{U}}  \cdot  \\ &\left(  \mathbf{\bar{U}} \cdot \nabla \mathbf{U}^{(0)} + \mathbf{U}^{(0)} \cdot \nabla \mathbf{\bar{U}}  + \mathbf{U}^{(0)} \cdot \nabla \mathbf{U}^{(0)}  \right)  \mathrm{dv}, \nonumber
\end{align}
where we have expanded our domain of integration from $V_2= \{ \mathbf{R} \in V : R \ge \xi\}$ to the entire empty channel $V_C$.  As we did for the inner integral, we can write the outer integral as an expansion in $\alpha$.
\begin{equation}
  \mathbf{F}_{L_2} = \rho U^2 \ell^2 ( \mathbf{k}_4 \alpha^4 + \mathbf{k}_5 \alpha^5 + \hdots \,) ~.
\end{equation}
Likewise, in order to calculate the terms $\mathbf{k}_4$ and $\mathbf{k}_5$, we sort the Stokes velocities by leading order in $\alpha$. Both even terms and odd terms from the method-of-reflections expansion contribute to the outer integral. In particular, since the even terms are computed numerically, the outer integral must also be computed numerically, rather than as a closed analytic formula.  

\subsection*{Inertial Lift Force}
The total lift force is the sum of the inner and outer integrals $\mathbf{F}_L = \mathbf{F}_{L_1} + \mathbf{F}_{L_2}$; combining the results from inner and outer expansions, we can then calculate the coefficients of the series expansion to obtain the following scaling law for the lift force
\begin{equation}
      \mathbf{F}_L(\bx_0) = \frac{\rho U^2 a^4}{H^2} \left[ \mathbf{c}_4(\bx_0) + \frac{a}{H} \mathbf{c}_5(\bx_0)\right] + O(a^6) \, .
\end{equation}
Recall that $\rho$ is the fluid density, $H$ is the channel height, $U$ is the centerline velocity of the undisturbed flow, $a$ is the particle radius, and $\mathbf{c}_4(\bx_0)$ and $\mathbf{c}_5(\bx_0)$ are dimensionless constants including both analytical and numerically computed components, and that depend on the location of the particle $\bx_0$ and the aspect ratio of the rectangular cross-section.  A text file with the values of $\mathbf{c}_4$ and $\mathbf{c}_5$, as well as the backflow correction $\mathbf{S}$ described in the main text for different particle locations is included in the ESI.

%%%END OF MAIN TEXT%%%


%\balance


%%%%REFERENCES%%%
\bibliographystyle{rsc}
%\bibliography{migration-velocity} 

\begin{mcitethebibliography}{32}
\providecommand*{\natexlab}[1]{#1}
\providecommand*{\mciteSetBstSublistMode}[1]{}
\providecommand*{\mciteSetBstMaxWidthForm}[2]{}
\providecommand*{\mciteBstWouldAddEndPuncttrue}
  {\def\EndOfBibitem{\unskip.}}
\providecommand*{\mciteBstWouldAddEndPunctfalse}
  {\let\EndOfBibitem\relax}
\providecommand*{\mciteSetBstMidEndSepPunct}[3]{}
\providecommand*{\mciteSetBstSublistLabelBeginEnd}[3]{}
\providecommand*{\EndOfBibitem}{}
\mciteSetBstSublistMode{f}
\mciteSetBstMaxWidthForm{subitem}
{(\emph{\alph{mcitesubitemcount}})}
\mciteSetBstSublistLabelBeginEnd{\mcitemaxwidthsubitemform\space}
{\relax}{\relax}

\bibitem[Amini \emph{et~al.}(2014)Amini, Lee, and Di~Carlo]{DiCarlo14}
H.~Amini, W.~Lee and D.~Di~Carlo, \emph{Lab Chip}, 2014, \textbf{14},
  2739--2761\relax
\mciteBstWouldAddEndPuncttrue
\mciteSetBstMidEndSepPunct{\mcitedefaultmidpunct}
{\mcitedefaultendpunct}{\mcitedefaultseppunct}\relax
\EndOfBibitem
\bibitem[Saffman(1965)]{Saffman65}
P.~G. Saffman, \emph{J. Fluid Mech.}, 1965, \textbf{22}, 385--400\relax
\mciteBstWouldAddEndPuncttrue
\mciteSetBstMidEndSepPunct{\mcitedefaultmidpunct}
{\mcitedefaultendpunct}{\mcitedefaultseppunct}\relax
\EndOfBibitem
\bibitem[Ho and Leal(1974)]{HoLeal74}
B.~P. Ho and L.~G. Leal, \emph{J. Fluid Mech.}, 1974, \textbf{65},
  365--400\relax
\mciteBstWouldAddEndPuncttrue
\mciteSetBstMidEndSepPunct{\mcitedefaultmidpunct}
{\mcitedefaultendpunct}{\mcitedefaultseppunct}\relax
\EndOfBibitem
\bibitem[Schonberg and Hinch(1989)]{Hinch89}
J.~A. Schonberg and E.~J. Hinch, \emph{J. Fluid Mech.}, 1989, \textbf{203},
  517--524\relax
\mciteBstWouldAddEndPuncttrue
\mciteSetBstMidEndSepPunct{\mcitedefaultmidpunct}
{\mcitedefaultendpunct}{\mcitedefaultseppunct}\relax
\EndOfBibitem
\bibitem[Hood \emph{et~al.}(2015)Hood, Lee, and Roper]{Hood15}
K.~Hood, S.~Lee and M.~Roper, \emph{J. Fluid Mech.}, 2015, \textbf{765},
  452--479\relax
\mciteBstWouldAddEndPuncttrue
\mciteSetBstMidEndSepPunct{\mcitedefaultmidpunct}
{\mcitedefaultendpunct}{\mcitedefaultseppunct}\relax
\EndOfBibitem
\bibitem[Di~Carlo \emph{et~al.}(2009)Di~Carlo, Edd, Humphry, Stone, and
  Toner]{DiCarlo09}
D.~Di~Carlo, J.~F. Edd, K.~J. Humphry, H.~A. Stone and M.~Toner, \emph{Phys.
  Rev. Lett.}, 2009, \textbf{102}, 094503\relax
\mciteBstWouldAddEndPuncttrue
\mciteSetBstMidEndSepPunct{\mcitedefaultmidpunct}
{\mcitedefaultendpunct}{\mcitedefaultseppunct}\relax
\EndOfBibitem
\bibitem[Prohm and Stark(2014)]{ProhmStark14}
C.~Prohm and H.~Stark, \emph{Lab Chip}, 2014, \textbf{14}, 2115--2123\relax
\mciteBstWouldAddEndPuncttrue
\mciteSetBstMidEndSepPunct{\mcitedefaultmidpunct}
{\mcitedefaultendpunct}{\mcitedefaultseppunct}\relax
\EndOfBibitem
\bibitem[Liu \emph{et~al.}(2015)Liu, Hu, Jiang, and Sun]{Liu2015}
C.~Liu, G.~Hu, X.~Jiang and J.~Sun, \emph{Lab Chip}, 2015, \textbf{15},
  1168--1177\relax
\mciteBstWouldAddEndPuncttrue
\mciteSetBstMidEndSepPunct{\mcitedefaultmidpunct}
{\mcitedefaultendpunct}{\mcitedefaultseppunct}\relax
\EndOfBibitem
\bibitem[Zhou and Papautsky(2013)]{Papautsky13}
J.~Zhou and I.~Papautsky, \emph{Lab Chip}, 2013, \textbf{13}, 1121--1132\relax
\mciteBstWouldAddEndPuncttrue
\mciteSetBstMidEndSepPunct{\mcitedefaultmidpunct}
{\mcitedefaultendpunct}{\mcitedefaultseppunct}\relax
\EndOfBibitem
\bibitem[Segr\'{e} and Silberberg(1961)]{SegreSilberberg61}
G.~Segr\'{e} and A.~Silberberg, \emph{Nature}, 1961, \textbf{189},
  209--210\relax
\mciteBstWouldAddEndPuncttrue
\mciteSetBstMidEndSepPunct{\mcitedefaultmidpunct}
{\mcitedefaultendpunct}{\mcitedefaultseppunct}\relax
\EndOfBibitem
\bibitem[Jeffrey and Pearson(1965)]{JeffreyPearson65}
R.~C. Jeffrey and J.~R.~A. Pearson, \emph{J. Fluid Mech.}, 1965, \textbf{22},
  721--735\relax
\mciteBstWouldAddEndPuncttrue
\mciteSetBstMidEndSepPunct{\mcitedefaultmidpunct}
{\mcitedefaultendpunct}{\mcitedefaultseppunct}\relax
\EndOfBibitem
\bibitem[Karnis \emph{et~al.}(1966)Karnis, Goldsmith, and Mason]{Karnis66}
A.~Karnis, H.~L. Goldsmith and S.~G. Mason, \emph{Can. J. Chem. Eng.}, 1966,
  \textbf{44}, 181--193\relax
\mciteBstWouldAddEndPuncttrue
\mciteSetBstMidEndSepPunct{\mcitedefaultmidpunct}
{\mcitedefaultendpunct}{\mcitedefaultseppunct}\relax
\EndOfBibitem
\bibitem[Tachibana(1973)]{Tachibana73}
M.~Tachibana, \emph{Rheol. Acta}, 1973, \textbf{12}, 58--69\relax
\mciteBstWouldAddEndPuncttrue
\mciteSetBstMidEndSepPunct{\mcitedefaultmidpunct}
{\mcitedefaultendpunct}{\mcitedefaultseppunct}\relax
\EndOfBibitem
\bibitem[Uijttewaal \emph{et~al.}(1994)Uijttewaal, Nijhof, and
  Heethaar]{Uijttewaal94}
W.~S. Uijttewaal, E.-J. Nijhof and R.~M. Heethaar, \emph{J. Biomech.}, 1994,
  \textbf{27}, 35 -- 42\relax
\mciteBstWouldAddEndPuncttrue
\mciteSetBstMidEndSepPunct{\mcitedefaultmidpunct}
{\mcitedefaultendpunct}{\mcitedefaultseppunct}\relax
\EndOfBibitem
\bibitem[Choi and Lee(2010)]{ChoiLee2010}
Y.-S. Choi and S.-J. Lee, \emph{Microfluid. Nanofluid.}, 2010, \textbf{9},
  819--829\relax
\mciteBstWouldAddEndPuncttrue
\mciteSetBstMidEndSepPunct{\mcitedefaultmidpunct}
{\mcitedefaultendpunct}{\mcitedefaultseppunct}\relax
\EndOfBibitem
\bibitem[Choi \emph{et~al.}(2011)Choi, Seo, and Lee]{ChoiSeoLee11}
Y.-S. Choi, K.-W. Seo and S.-J. Lee, \emph{Lab chip}, 2011, \textbf{11},
  460--465\relax
\mciteBstWouldAddEndPuncttrue
\mciteSetBstMidEndSepPunct{\mcitedefaultmidpunct}
{\mcitedefaultendpunct}{\mcitedefaultseppunct}\relax
\EndOfBibitem
\bibitem[Gossett \emph{et~al.}(2012)Gossett, Tse, Dudani, Goda, Woods, Graves,
  and Di~Carlo]{SMLL:DiCarlo2012}
D.~R. Gossett, H.~T.~K. Tse, J.~S. Dudani, K.~Goda, T.~A. Woods, S.~W. Graves
  and D.~Di~Carlo, \emph{Small}, 2012, \textbf{8}, 2757--2764\relax
\mciteBstWouldAddEndPuncttrue
\mciteSetBstMidEndSepPunct{\mcitedefaultmidpunct}
{\mcitedefaultendpunct}{\mcitedefaultseppunct}\relax
\EndOfBibitem
\bibitem[Asmolov(1999)]{Asmolov99}
E.~S. Asmolov, \emph{Journal of Fluid Mechanics}, 1999, \textbf{381},
  63--87\relax
\mciteBstWouldAddEndPuncttrue
\mciteSetBstMidEndSepPunct{\mcitedefaultmidpunct}
{\mcitedefaultendpunct}{\mcitedefaultseppunct}\relax
\EndOfBibitem
\bibitem[Sheng \emph{et~al.}(2008)Sheng, Malkiel, and Katz]{sheng2008using}
J.~Sheng, E.~Malkiel and J.~Katz, \emph{Experiments in fluids}, 2008,
  \textbf{45}, 1023--1035\relax
\mciteBstWouldAddEndPuncttrue
\mciteSetBstMidEndSepPunct{\mcitedefaultmidpunct}
{\mcitedefaultendpunct}{\mcitedefaultseppunct}\relax
\EndOfBibitem
\bibitem[Katz and Sheng(2010)]{katz2010applications}
J.~Katz and J.~Sheng, \emph{Annual Review of Fluid Mechanics}, 2010,
  \textbf{42}, 531--555\relax
\mciteBstWouldAddEndPuncttrue
\mciteSetBstMidEndSepPunct{\mcitedefaultmidpunct}
{\mcitedefaultendpunct}{\mcitedefaultseppunct}\relax
\EndOfBibitem
\bibitem[Choi \emph{et~al.}(2012)Choi, Seo, Sohn, and Lee]{choi2012advances}
Y.-S. Choi, K.-W. Seo, M.-H. Sohn and S.-J. Lee, \emph{Optics and Lasers in
  Engineering}, 2012, \textbf{50}, 39--45\relax
\mciteBstWouldAddEndPuncttrue
\mciteSetBstMidEndSepPunct{\mcitedefaultmidpunct}
{\mcitedefaultendpunct}{\mcitedefaultseppunct}\relax
\EndOfBibitem
\bibitem[Duffy \emph{et~al.}(1998)Duffy, McDonald, Schueller, and
  Whitesides]{Whitesides98}
D.~C. Duffy, J.~C. McDonald, O.~J.~A. Schueller and G.~M. Whitesides,
  \emph{Anal. Chem.}, 1998, \textbf{70}, 4974--4984\relax
\mciteBstWouldAddEndPuncttrue
\mciteSetBstMidEndSepPunct{\mcitedefaultmidpunct}
{\mcitedefaultendpunct}{\mcitedefaultseppunct}\relax
\EndOfBibitem
\bibitem[Roper \emph{et~al.}(2013)Roper, Simonin, Hickey, Leeder, and
  Glass]{Roper13}
M.~Roper, A.~Simonin, P.~C. Hickey, A.~Leeder and N.~L. Glass, \emph{Proc. Nat.
  Acad. Sci. U.S.A.}, 2013, \textbf{110}, 12875--12880\relax
\mciteBstWouldAddEndPuncttrue
\mciteSetBstMidEndSepPunct{\mcitedefaultmidpunct}
{\mcitedefaultendpunct}{\mcitedefaultseppunct}\relax
\EndOfBibitem
\bibitem[Sveen(2004)]{matpiv}
J.~K. Sveen, \emph{Mech. Appl. Math.}, 2004\relax
\mciteBstWouldAddEndPuncttrue
\mciteSetBstMidEndSepPunct{\mcitedefaultmidpunct}
{\mcitedefaultendpunct}{\mcitedefaultseppunct}\relax
\EndOfBibitem
\bibitem[Leal(1980)]{Leal80}
L.~G. Leal, \emph{Annu. Rev. Fluid Mech.}, 1980, \textbf{12}, 435--476\relax
\mciteBstWouldAddEndPuncttrue
\mciteSetBstMidEndSepPunct{\mcitedefaultmidpunct}
{\mcitedefaultendpunct}{\mcitedefaultseppunct}\relax
\EndOfBibitem
\bibitem[Happel and Brenner(1982)]{HappelBrenner83}
J.~Happel and H.~Brenner, \emph{Low Reynolds number hydrodynamics: with special
  applications to particulate media}, Springer, 1982, vol.~1\relax
\mciteBstWouldAddEndPuncttrue
\mciteSetBstMidEndSepPunct{\mcitedefaultmidpunct}
{\mcitedefaultendpunct}{\mcitedefaultseppunct}\relax
\EndOfBibitem
\bibitem[Sollier \emph{et~al.}(2011)Sollier, Murray, Maoddi, and
  Di~Carlo]{sollier2011rapid}
E.~Sollier, C.~Murray, P.~Maoddi and D.~Di~Carlo, \emph{Lab Chip}, 2011,
  \textbf{11}, 3752--3765\relax
\mciteBstWouldAddEndPuncttrue
\mciteSetBstMidEndSepPunct{\mcitedefaultmidpunct}
{\mcitedefaultendpunct}{\mcitedefaultseppunct}\relax
\EndOfBibitem
\bibitem[Ciftlik \emph{et~al.}(2013)Ciftlik, Ettori, and Gijs]{Gijs2013}
A.~T. Ciftlik, M.~Ettori and M.~A.~M. Gijs, \emph{Small}, 2013, \textbf{9},
  2764--2773\relax
\mciteBstWouldAddEndPuncttrue
\mciteSetBstMidEndSepPunct{\mcitedefaultmidpunct}
{\mcitedefaultendpunct}{\mcitedefaultseppunct}\relax
\EndOfBibitem
\bibitem[Miura \emph{et~al.}(2014)Miura, Itano, and
  Sugihara-Seki]{Sugihara-Seki2014}
K.~Miura, T.~Itano and M.~Sugihara-Seki, \emph{J. Fluid Mech.}, 2014,
  \textbf{749}, 320--330\relax
\mciteBstWouldAddEndPuncttrue
\mciteSetBstMidEndSepPunct{\mcitedefaultmidpunct}
{\mcitedefaultendpunct}{\mcitedefaultseppunct}\relax
\EndOfBibitem
\bibitem[Papanastasiou \emph{et~al.}(1999)Papanastasiou, Georgiou, and
  Alexandrou]{Papanastasiou99}
T.~C. Papanastasiou, G.~C. Georgiou and A.~N. Alexandrou, \emph{Viscous Fluid
  Flow}, CRC Press, 1999\relax
\mciteBstWouldAddEndPuncttrue
\mciteSetBstMidEndSepPunct{\mcitedefaultmidpunct}
{\mcitedefaultendpunct}{\mcitedefaultseppunct}\relax
\EndOfBibitem
\bibitem[Lamb(1945)]{Lamb45}
H.~Lamb, \emph{Hydrodynamics}, Dover Publications, 1945\relax
\mciteBstWouldAddEndPuncttrue
\mciteSetBstMidEndSepPunct{\mcitedefaultmidpunct}
{\mcitedefaultendpunct}{\mcitedefaultseppunct}\relax
\EndOfBibitem
\bibitem[Kim and Karrila(2005)]{KimKarrila2005}
S.~Kim and S.~Karrila, \emph{Microhydrodynamics: Principles and Selected
  Applications}, Dover Publications, 2005\relax
\mciteBstWouldAddEndPuncttrue
\mciteSetBstMidEndSepPunct{\mcitedefaultmidpunct}
{\mcitedefaultendpunct}{\mcitedefaultseppunct}\relax
\EndOfBibitem
\end{mcitethebibliography}
% %the RSC's .bst file

\providecommand{\noopsort}[1]{}\providecommand{\singleletter}[1]{#1}%
\providecommand*{\mcitethebibliography}{\thebibliography}
\csname @ifundefined\endcsname{endmcitethebibliography}
{\let\endmcitethebibliography\endthebibliography}{}

\end{document}